\documentclass[prl,aps,twocolumn,superscriptaddress,notitlepage,10pt
]{revtex4-2}
\usepackage[T1]{fontenc}
\usepackage{amssymb,amsfonts,amsmath,amstext,graphicx}
\usepackage{amsmath}
\usepackage{tikz}
\usepackage{graphicx}
\usepackage{svg}
\usepackage[normalem]{ulem}
\usepackage{tabularx}
\def\bra#1{\langle{#1}|}
\def\ket#1{|{#1}\rangle}
\def\braket#1{\langle{#1}\rangle}
\def\Bra#1{\left\langle#1\right|}
\def\Ket#1{\left|#1\right \rangle}
\def\BraVert{\egroup\,\mid\,\bgroup}
\def\lket#1{\vert#1\rangle\hspace{-1mm}\rangle}
\def\lbra#1{\langle\hspace{-1mm}\langle#1\vert}
\def\Brak#1#2#3{\bra{#1}#2\ket{#3}}
\usepackage{braket}
\usepackage{xcolor}
\usepackage{ marvosym }
\usepackage[colorlinks, linkcolor=red,urlcolor=blue,citecolor=blue]{hyperref}

\usepackage{algorithm}
\usepackage{algpseudocode}
\usepackage{float}
\usepackage{graphicx}
\usepackage{dcolumn}
\usepackage{bm}
\begin{document}
\title{Nonequilibrium steady states induced by stochastic mid-circuit measurements and resets on a quantum computer} 
\author{Jakob Murauer}
\affiliation{Research Institute CODE, University of the Bundeswehr Munich, 81739 Munich, Germany}\author{Sabine Tornow} \affiliation{Research Institute CODE, University of the Bundeswehr Munich, 81739 Munich, Germany}\author{Gabriele Perfetto}
\affiliation{Institut für Theoretische Physik, ETH Zürich, Wolfgang-Pauli-Str. 27, 8093 Zürich, Switzerland}

\begin{abstract}
Stochastic resetting has emerged as a versatile protocol to drive quantum many-body systems to non-equilibrium steady states by interspersing unitary dynamics with measurements and resets at random times. In spite of this, a quantum hardware validation of such non-equilibrium steady states is still missing. Here, we achieve this goal by first formulating a noisy discrete-time theory where unitary gates alternate with noisy mid-circuit projective measurements and conditional resets. This noisy conditional resetting theory is then 
demonstrated on a superconducting quantum processor for up to $N=7$ qubits. We consider, as a paradigmatic case, the unitary dynamics of the interacting Floquet transverse-field Ising model. The stationary state of the noisy conditional resetting agrees quantitatively with the experiments, and it shows crossover behavior related to the equilibrium quantum phase transition of the model. Our results might thus pave the way for the preparation of collective stationary states on noisy quantum devices and for further developments of quantum algorithms involving mid-circuit measurements. 
\end{abstract}
\maketitle

\paragraph{Introduction.---} Stochastic resetting describes a very intuitive idea: a process is interrupted at random times and starts over from its initial configuration \cite{evans2011resetting,evans2011resetlong}. As such, this process is ubiquitous in nature and found applications across multiple areas of science such as algorithm optimization \cite{tong2008random,lorenz2017runtime,blumer2024combining,church2025accelerating}, accelerating search processes \cite{renewal1,renewal2,renewal3,Campos2015,renewal5,Sandev_1,Sandev_book} and generating non-equilibrium steady states (NESS) displaying correlations \cite{evans2014diffarbitraryd,DPTreset2015,eule2016non,power_law_reset_ness,ness_reset_random_walk,renewal8,reset_geometric_brownian}. The latter has recently found experimental verification in colloidal particles, where stochastic resets are implemented by switching the confining optical traps associated to each particle \cite{resetting_colloids_1,resetting_colloids_2}.

In quantum systems resetting amounts to projecting the quantum state of the system of interest to a reference reset state \cite{Hartmann2006,Linden2010,Rose2018spectral,Mukherjee2018,carollo2019unravelling,Armin2020,riera2020measurement,quantum_reset_ldev}, similarly to a global projective measurement. The joint effect of resets and projective measurements has been, indeed, theoretically analyzed, suggesting a number of fruitful applications in quantum technologies. These include the speed-up of the mean hitting time of quantum walks \cite{dhar2015,dhar2015b,grunbaum2013recurrence,Barkai2016,Barkai2017,Barkai2018,tornow2023,yin2025resonance,Walter25} subject to projective measurements at stroboscopic times \cite{majumdar2023,restart_Yin,Modak_resetting,roy2025causality}, and accelerated state preparation in dissipative quantum systems \cite{bao2022acceleratingrelaxationmarkovianopen,solanki2025universal}. Moreover, conditional resetting protocols, where a global projective measurement determines the subsequent reset state, have been shown to generate NESS displaying collective behavior \cite{perfetto2021designing,turkeshireset,Magoni_PRA_quantum,dattagupta2022stochastic,sevilla2023dynamics,Anish_conditional,wald2025stochastic,kulkarni2025dynamically}. Such NESS can be exploited for enhanced sensing in quantum metrology \cite{sensing_1,sensing_2,sensing_3,sensing_4,sensing_5}. In spite of this importance, stochastic-resetting-induced non-equilibrium steady states in interacting quantum many-body systems have, to the best of our knowledge, not yet been demonstrated on quantum hardware. Their realization is challenging because mid-circuit readout errors, hardware noise, and decoherence can corrupt the reset operation and alter the resulting stationary state.

\begin{figure}[!t]
    \centering   \includegraphics[width=\columnwidth]{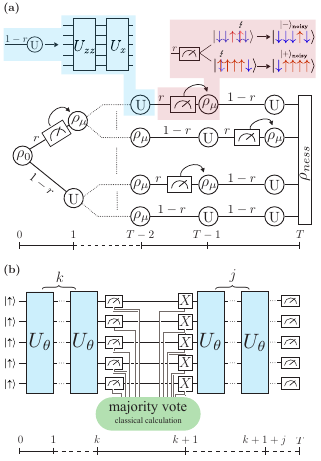}
    \caption{\textit{Noisy reset protocol and its hardware implementation}. (a) For poissonian resetting, at each discrete time step, the system either evolves unitarily with probability $1-r$ (the unitary is sketched in the blue panel), or undergoes a reset. In the conditional protocol, the reset state is determined by a majority vote on the mid-circuit measurement (oscilloscope drawings) outcomes. We account in the theory for readout errors, which effectively result into a depolarization of the reset state (red panel). Averaging over many stochastic realizations of resetting yields a NESS at long times. (b) Quantum circuit implementation of a single reset realization. Unitary gates $U_{\theta}$ are applied until the first reset is performed (in the figure at the $k$th step). After a mid-circuit measurement is executed, real-time classical operations perform the majority vote. Based on the result, single-qubit rotations ($X$ gates) are conditionally applied to the qubits to align the minority of spins with the measured majority. This protocol is repeated until a final time $T$, when a readout is performed.}
\label{fig:overview}
\end{figure}
Here, we realize the NESS generated by stochastic resets and mid-circuit measurements in remote experiments on a superconducting quantum processor for systems of up to $N=7$ qubits, see Fig.~\ref{fig:overview}. To do this we formulate a noisy conditional resetting theory in discrete time, where unitary gates are interspersed with noisy mid-circuit readouts, cf. Fig.~\ref{fig:overview}(a). We consider as a paradigmatic testbed the interacting Floquet transverse-field Ising model, which shows a quantum phase transition associated to $\mathbb{Z}_2$ symmetry breaking at zero temperature. Noise in readout errors is accounted for in the theory through a $\mathbb{Z}_2$-symmetry inspired noise model, where quantum many-body states are grouped together depending on the number of bitflips with respect to the background ordered reset states. On the quantum hardware our implementation exploits dynamic circuits \cite{rqh3,rqh4,rqh5}, where a sequence of unitary spin-flip operations are applied after processing the classical information concurrent with the readout, see Fig.~\ref{fig:overview}(b). The NESS predicted by the noisy theory is in excellent agreement with the experimental data and it features crossover behavior reminiscent of the quantum phase transition of the model. Our analysis thus shows that collective stationary states arising from interacting many-body quantum dynamics and resets can be faithfully implemented on state-of-the-art quantum processors. In addition, our results provide practical insights for further development of quantum algorithms based on dynamic circuits. 

\paragraph{Resetting protocols---} 
The simplest implementation of stochastic resetting is achieved by reinitializing the many-body state of the system to a prescribed reset state, denoted as $\ket{0}$, at random times. We name this protocol henceforth \textit{unconditional resetting}. On digital quantum simulators this protocol is implemented in discrete time, as sketched in Fig.~\ref{fig:overview}(a). This means that at each discrete time step $n \in \mathbb{N}_0$ the system is either projected with probability $0\leq r_n \leq 1$ to the reset state, or a unitary gate $U$ is applied with probability $1-r_n$. Here, $r_n$ crucially depends on the number $n$ of steps elapsed since the last reset. The dynamics therefore does not carry memory of what happened prior to a resetting event, which identifies the process as renewal \cite{evans2020review,nagar2023review,feller1957introduction}. The density matrix $\rho(t)$ at discrete time $t$ is a mixed state obtained by averaging over all stochastic realizations of the resetting process
\begin{equation}
\rho(t)=\sum_{n=0}^{t} P_n(t)\ket{n}\bra{n}, \quad \mbox{with} \quad \rho(0)=\ket{0}\bra{0}.
\label{eq:density_matrix}
\end{equation}
with $P_n(t)$ denoting the probability of staying in state $\ket{n}=U^n \ket{0}$ at time $t$ without knowledge on when the previous reset took place. Reset projections are nonunitary operations and drive the system at long times to a non-equilibrium steady state 
(NESS) density matrix $\rho^{\mathrm{ness}}$. The derivation of the latter is detailed in the Supplemental Material \cite{SM}. The final expression is
\begin{equation}
\rho^{\mathrm{ness}}= P_0^{\mathrm{stat}} \sum_{n=1}^{\infty}q_n \, \ket{n}\bra{n} + P_0^{\mathrm{stat}}\ket{0}\bra{0},
\label{eq:stationary_state_unc}
\end{equation}
where $q_n=\prod_{j=0}^{n-1}(1-r_j)$ is the survival probability, i.e., the probability that no reset is performed after $n$ steps since the last reset event. Survival probability also determines the stationary probability $P_0^{\mathrm{stat}}$ of performing a reset $P_0^{\mathrm{stat}}=1/(1+\sum_{n=0}^{\infty}q_{n+1})$. We consider henceforth poissonian resetting, where the resetting probability $r_n=r$ is constant for all $n$. In this case, a NESS is reached and Eq.~\eqref{eq:stationary_state_unc} matches the result of Ref.~\cite{wald2025stochastic}. 

We then move on by considering \textit{conditional resetting}. In this protocol, the reset state is chosen within a set of states. We consider here two reset states for a system of $N$ qubits: $\ket{+}=\ket{\uparrow_1,\dots \uparrow_N}$ and $\ket{-}=\ket{\downarrow_1,\dots \downarrow_N}$. These states are thus completely magnetized in the positive and negative $z$ direction of the Pauli basis. The unitary evolution from the reset states is denoted as $\ket{n_{\pm}}= U^n \ket{\pm}$.  The choice of the reset state is done on the basis of a global projective measurement. This is expressed by the transition matrix 
\begin{equation}
P_{i,j}(n)=\braket{n_i| \hat{P}_j|n_i}, \quad \mbox{with} \quad i,j\in \{\pm\},
\label{eq:P_matrix}
\end{equation}
which gives the probability that, after $n$ steps since the previous reset to state $\ket{i}$, the system is reset to state $\ket{j}$. The operator $\hat{P}_j$ projects onto the subspace associated with measurement outcome $j$, and thus defines the measurement protocol. Here, we adopt a majority protocol~\cite{perfetto2021designing,Anish_conditional,soldner2025nonanaliticities}, where $\ket{+}$ or $\ket{-}$ is selected depending on whether the majority of measured spins points in the up or down direction, respectively. This protocol is illustrated in Fig.~\ref{fig:overview}(a). It defines a semi-Markov process \cite{janssen2006applied}, where the choice of the reset state $j$ does not only depend on the previous reset state $i$, but also on the stochastic discrete time $n$ elapsed according to the transition matrix $R_{i,j}(n)=P_{i,j}(n)r_n$. The derivation of the ensuing NESS density matrix is detailed in the End Matter and it reads as 
\begin{equation}
\rho^{\mathrm{ness}}=\sum_{j=\pm}\left( P^{\mathrm{stat}}_{j}\,\sum_{n=1}^{\infty}q_n \ket{n_j}\bra{n_j} +P^{\mathrm{stat}}_{j}\ket{j}\bra{j}\right).
\label{eq:ness_cond}
\end{equation}
The stationary probabilities $P_{\pm}^{\mathrm{stat}}$ of resetting to state $\ket{\pm}$ are given by 
\begin{equation}
P_{\pm}^{\mathrm{stat}}=\frac{\hat{g}_{\mp,\pm}(1)}{[1+\sum_{n=0}^{\infty}q_{n+1}][\hat{g}_{\mp,\pm}(1)+\hat{g}_{\pm,\mp}(1) ]},
\label{eq:stationary_prob_cond}
\end{equation}
with the definition $\hat{g}_{i,j}(z)=\sum_{n=1}^{\infty}z^{-n} q_n R_{i,j}(n)$. Equation \eqref{eq:ness_cond} expresses the NESS from conditional resetting as a statistical mixture of the evolution corresponding to the states $\ket{\pm}$. The weight of each of the two terms is given by the stationary resetting probabilities \eqref{eq:stationary_prob_cond} and it therefore depends on the measurement protocol.

\paragraph{Ising gates and order parameter---} For the unitary dynamics, we consider the interacting Floquet transverse-field Ising model (FTFIM) \cite{prosen2007chaos,prosen2000exact,bertini_FTIM} whose unitary gate $U_{\theta}$ is parametrized by the angle $\theta$ as
\begin{equation}
U_{\theta}=U_x U_{zz}, \quad \mbox{with} \quad U_{x}=e^{-i H_x \theta}, \quad U_{zz}=e^{-i H_{zz}\theta}.
\label{eq:ising_discrete_unitary}
\end{equation}
and 
\begin{equation}
H_x=-J h\sum_{i=1}^{N}\sigma^x_i, \quad \mbox{and} \quad H_{zz}=-J\sum_{i=1}^{N}\sigma^z_i \sigma^z_{i+1}.
\label{eq:ising_discrete_terms}
\end{equation}
Here, $J>0$ is the ferromagnetic coupling, $h$ the transverse field, and $\sigma_i^{\alpha}$ the Pauli operator acting on site $i$ in direction $\alpha\in{x,y,z}$. The Floquet eigenspectrum becomes gapless at $h=1$ \cite{FTIM_1,FTIM_2,FTIM_3,SM} in the thermodynamic limit $N\to \infty$. This reflects the quantum phase transition taking place for the TFIM Hamiltonian, with spontaneous $\mathbb{Z}_2$ symmetry breaking. The order parameter for the phase transition of the TFIM is the longitudinal magnetization density $m=\sum_{i=1}^N \sigma^z_i/N$, which acquires a nonzero value in the ferromagnetic phase ($h<1$). For unconditional resetting, we thus study the corresponding NESS value $m_{\mathrm{ness}}^{\mathrm{uncond.}}=\mbox{Tr}[m \rho^{\mathrm{ness}}]$ according to Eq.~\eqref{eq:stationary_state_unc}. For conditional resetting, the $\mathbb{Z}_2$ symmetry implies $\hat{g}_{+,-}(1)=\hat{g}_{-,+}(1)$ in Eq.~\eqref{eq:stationary_prob_cond}, yielding a vanishing NESS magnetization, $\langle m\rangle_{\mathrm{ness}}^{\mathrm{cond.}}=0$. We therefore focus on the squared magnetization density, whose NESS expectation value is given by
\begin{equation}
\braket{m^2}_{\mathrm{ness}}^{\mathrm{cond.}}=r \sum_{t=0}^{\infty}(1-r)^t \braket{t_+|m_z^2|t_+}.
\label{eq:final_m2_conditional_poisson}
\end{equation}
We move on by describing the experimental implementation of the resetting dynamics on the quantum processor.

\paragraph{Quantum hardware implementation.---}
Experiments have been remotely conducted on the IBM Quantum \texttt{ibm\_marrakesh} processor. We initialize the system in the state $\ket{+}=\ket{
\uparrow}^{\otimes N}$ for the $N$ qubits. We implement the unitary evolution in Eqs.~\eqref{eq:ising_discrete_unitary} and \eqref{eq:ising_discrete_terms} as a sequence of dynamic quantum circuits using Qiskit \cite{javadiabhari2024quantumcomputingqiskit}. To realize the $ZZ$ couplings, we employ a standard gate decomposition of $U_{zz}$ in terms of two-qubits $\mathrm{CNOT}_{i\to i+1}$ and single-qubit rotation $RZ_{i}(2\theta)$ gates \cite{nielsen2010quantum,SM}; The transverse-field unitary is implemented as a series of single-qubit rotations $RX_i(2\theta)$. Since we consider periodic boundary conditions, the limited native connectivity of the heavy-hex hardware lattice is insufficient. We therefore introduce SWAP gates to route the necessary non-local interactions. The experimental realization of the reset protocols differs fundamentally based on the conditioning.
 
For the unconditional reset protocol, we circumvent active hardware resets by exploiting the renewal structure of the dynamics. We thus implement system dynamics through pure unitary evolution corresponding to the time elapsed since the last stochastically assigned reset event. Because this approach does not require active mid-circuit measurement, we are able to extensively apply error mitigation techniques, including dynamical decoupling \cite{DD1, DD2}, readout error-mitigation \cite{TREX1, TREX2}, zero-noise extrapolation \cite{ZNE1, ZNE2} and Pauli twirling \cite{PhysRevLett.82.2417DD, PhysRevLett.119.180509ZNE, PhysRevA.94.052325twirling, PhysRevA.105.032620TREX}.

In contrast, the conditional reset protocol requires active mid-circuit interventions, since the dynamics \eqref{eq:P_matrix} implements a semi-Markov process. We perform intermediate measurements in the $Z$-basis and evaluate a classical majority-vote function in real time \cite{SM}. Based on this classical processing of the measurement output, conditional single-qubits $RX_i(\pi)$ gates are dynamically applied to the qubits to prepare the target reset state, as sketched in Fig.~\ref{fig:overview}(b). Due to control hardware limitations associated with dynamic circuit execution, advanced error mitigation could not be integrated into the conditional reset experiments. Quantum mechanical expectation values are eventually computed by averaging over a number $M=10^4$ of readouts (in the Pauli $z$ direction) at the end of the circuit for each reset realization.
\begin{figure*}[!t]
    \centering
    \includegraphics[width=\textwidth]{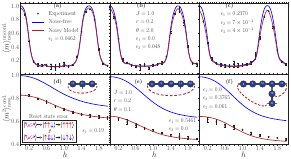}
    \caption{\textit{Comparison between experimental results and theoretical predictions of unconditional and conditional reset protocols.} (a)-(c) Unconditional reset protocol for 3, 5, and 7 qubits, respectively, evaluated up to a maximum time $T=10^5$. (d)-(f) Conditional reset protocol for 3, 5, and 7 qubits, evaluated up to $T=400$. In all panels, experimental hardware results (dots) are compared with noiseless theoretical simulations (blue solid lines) and theoretical predictions from the noisy reset model (red solid lines), sketched in panel (d). Error bars denote $95\%$ confidence intervals derived from 5 independent experimental runs. The optimal fitted noise probabilities $\epsilon_i$ are reported in the figure together with system parameters. Averages are taken over $10^3$
  quantum trajectories for all experiments. Insets in (d)-(f) illustrate the hardware qubit topologies utilized for both the protocols; dashed red lines indicate non-native hardware couplings implemented via SWAP gates.}
    \label{fig:experimental_comparison}
\end{figure*}
\paragraph{Noise model.---}
In Fig.~\ref{fig:experimental_comparison}(a)-(c), we present the comparison between quantum hardware experiments (black dots) and unconditional resetting in Eq.~\eqref{eq:stationary_state_unc} (blue solid line). In Fig.~\ref{fig:experimental_comparison}(d)-(f) the comparison between conditional resetting \eqref{eq:final_m2_conditional_poisson} and hardware is shown. In both the cases, we find that the experimental data experience a downshift compared to the theory model. This downshift is more significantly pronounced for the conditional protocol. We now introduce a noise model to explain and quantify these discrepancies. 
 Reset operations on current quantum hardware are imperfect \cite{rost_long-time_2025}. We therefore account for bit-flip errors during the preparation of the reset state by modeling the reset as a noisy initialization process with finite fidelity. In the conditional resetting protocol, readout errors, as sketched in Fig.~\ref{fig:overview}(a) and \ref{fig:experimental_comparison}(d), during mid-circuit measurements can lead to the erroneous application of conditional bitflip gates. This effectively disorders the reset state compared to the noiseless resets states $\ket{\pm}$ that are completely ferromagnetically ordered in the positive/negative direction. We describe this phenomenon in the theory by using a mixed reset state $\rho_r$ rather than the pure states $\ket{\pm}$. We write the state $\rho_r$ as
\begin{equation}
\rho_r = \sum_{n_{bf}=0}^{N} \sum_{\mu} p_{n_{bf}}^{\mu} |n_{bf}^{\mu}\rangle\langle n_{bf}^{\mu}|.
\label{eq:mixed_reset_state}
\end{equation}
Here $|n_{bf}^{\mu}\rangle$ represents the $\mu$-th permutation of a state in the computational $Z$ basis with a number $n_{bf}$ of bitflips with respect to the reference state $\ket{+}$ \footnote{Because of the $\mathbb{Z}_2$ symmetry, the same analysis holds with $\ket{-}$ as reference state}, and $p_{n_{bf}}^{\mu}$ is the corresponding weight in the statistical mixture. Translation invariance enforces equal weight for all states with the same number of bit flips, so that $p^{\mu}_{n_{\mathrm{bf}}}=p_{n_{\mathrm{bf}}}/\Omega(n_{\mathrm{bf}})$, where $\Omega(n_{\mathrm{bf}})$ denotes the number of configurations with $n_{\mathrm{bf}}$ bit flips. Furthermore, the $\mathbb{Z}_2$ symmetry of the gates in Eqs.~\eqref{eq:ising_discrete_unitary} and \eqref{eq:ising_discrete_terms} implies that states with $n_{\mathrm{bf}}$ bit flips contribute equally to the expectation value as states with $N-n_{\mathrm{bf}}$ bit flips. By grouping the symmetry-paired probabilities into combined error parameters $\epsilon_k=p_k+p_{N-k}$, we can write a compact expression for the magnetization expectation value:
\begin{align}
\text{Tr}[(U_\theta^t)^\dagger m_z^2 U_\theta^t \rho_r] &= \left(1 - \sum_{k=1}^{(N-1)/2} \epsilon_k\right) \langle + | (U_\theta^t)^\dagger m_z^2 U_\theta^t | + \rangle \nonumber \\
&+\sum_{k=1}^{(N-1)/2} \epsilon_k \text{Tr}[(U_\theta^t)^\dagger m_z^2 U_\theta^t \rho_k],
\label{eq:depolarization_unitary}
\end{align}
where $\rho_{n_{bf}} =  \sum_{\mu=1}^{\Omega(n_{bf})} |n_{bf}^{\mu}\rangle\langle n_{bf}^{\mu}|/\Omega(n_{bf})$. The physical interpretation is that the system evolves from the ideal reset state $|+\rangle$ with probability $1 - \sum \epsilon_k$, and from translationally invariant depolarized $k$-bitflip states with probability $\epsilon_k$, reducing the overall magnetization. The NESS squared magnetization density for the noisy reset state is computed as
\begin{equation}
\langle m^2 \rangle^{\mathrm{cond.}}_{ness} = r \sum_{t=0}^{\infty} (1-r)^t \text{Tr}[(U_\theta^t)^\dagger m_z^2 U_\theta^t \rho_r].
\label{eq:noisy_m2}
\end{equation}
together with Eq.~\eqref{eq:depolarization_unitary}. This symmetry-reduced approach enables the numerical evaluation of NESS expectation values with only a polynomial number $(N-1)/2$ in the system size of fitting parameters and minimal computational overhead.

\paragraph{Unconditional resetting.---}
In Figs.~\ref{fig:experimental_comparison}(a)-(c), we compare the experimental data with the noisy reset theory (red solid lines). Due to the large $\theta=2$ value, one can observe inherent features of the digital quantum dynamics compared to the continuum-time limit. For $h>\pi/4$, the stroboscopic Floquet dynamics gradually restores ferromagnetic order, which reaches unity at $h=\pi/2$. For the latter transverse field value, the unitary gate \eqref{eq:ising_discrete_unitary} and \eqref{eq:ising_discrete_terms} acts, indeed, as the identity on both states $\ket{\pm}$ and therefore ferromagnetic ordering of such states cannot be lost. In addition, local maxima of the magnetization density at $h\approx 0.56$ and $h \approx 1.01$ are present for $N=3$ qubits in Fig.~\ref{fig:experimental_comparison}(a). These are caused by the degeneracy of a pair of Floquet quasienergies \cite{SM}. For larger system sizes, in Fig.~\ref{fig:experimental_comparison}(b) and (c), the Floquet spectrum becomes denser and the effect of such degeneracies consequently decreases. We also note that for unconditional resetting the effect of noisy readouts is weak and the agreement is excellent for all system sizes. We attribute the high fidelity of these experimental results to two primary factors. First, the data benefits from the application of advanced error mitigation techniques. Second, the unconditional reset protocol avoids the implementation of dynamic circuits and mid-circuit measurements by exploiting the renewal structure of resetting.

\paragraph{Conditional resetting.---} In Fig.~\ref{fig:experimental_comparison}(d)-(f), we turn to the conditional reset protocol. In this case the small $\theta=0.1$ value allows to approach the continuum-time limit. Ferromagnetic ordering quantified by the order parameter $\braket{m^2}_{\mathrm{ness}}^{\mathrm{cond.}}$ is, indeed, monotonically lost as the strength of the quantum fluctuations mediated by the transverse field $h$ increases. The quantum critical behavior of the TFIM is thus reflected in a smooth NESS crossover around $h=1$ on the quantum hardware. We observe an increased discrepancy between the noisy theoretical model \eqref{eq:noisy_m2} and the hardware results for larger system sizes $N=7$. This follows from the fact that classical evaluations according to the majority rule and the dynamical application of feedback gates must be completed entirely within the stringent coherence time limits of the superconducting qubits. The unavoidable latency, measurement cross-talk, and decoherence introduced by these active, mid-circuit, interventions act as strong sources of hardware noise \cite{PhysRevA.109.062617}. A quantitative comparison with alternative noise models, including depolarizing, phase-damping, generalized amplitude-damping, and correlated $ZZ$ noise, is provided in the Supplemental Material \cite{SM}. Among these models, the reset-state error model reliably yields a very good fit to the experimental data.
\paragraph{Conclusion.---} We have realized a stochastic-resetting-induced non-equilibrium steady state on superconducting quantum hardware. The protocol combines unitary dynamics with mid-circuit measurements and conditional resets. To compare with the quantum hardware data, we develop a theoretical model that accounts for noise in the reset operations. For the interacting Floquet transverse-field Ising model, the predictions from the noisy resetting theory quantitatively match the hardware experiments for systems of up to $N=7$ qubits. Our results suggest that stochastic resetting with dynamic circuits can provide a practical route toward engineering collective stationary states on noisy quantum processors. In the conditional protocol, this is achieved through mid-circuit measurements and real-time majority-vote logic. This connects stochastic resetting to the broader use of classical feed-forward as a resource for state preparation on near-term quantum machines~\cite{rqh5}. Quantum simulations for large system sizes are still challenging. We thus plan to integrate more sophisticated quantum error mitigation protocols \cite{PhysRevA.109.062617} into the conditional feedback loops. This analysis could eventually enable a systematic scaling to larger system sizes of quantum simulation of resetting-induced collective stationary states.
\paragraph{Data availability statement}
The data supporting the findings of this article can be accessed on Zenodo \cite{ZENODO_CITATION}.
\paragraph{Acknowledgements.---}
We acknowledge the use of IBM Quantum services for this work. The views expressed are those of the authors, and do not reflect the official policy or position of IBM or the IBM Quantum team.
In this manuscript, we used the \texttt{ibm\_marrakesh} quantum processor, which is one of IBM's 156-qubit Heron r2 superconducting transmon devices with heavy-hex connectivity. S.T. gratefully acknowledges insightful discussions with Eli Barkai. G.P. acknowledges funding from the Swiss National Science Foundation (SNSF) through Grant No.~10005336.


\bibliography{bibliography}
\newpage
\onecolumngrid
\section{End Matter}

\subsection{Derivation of the NESS density matrix for conditional resetting}
We here detail the derivation of the NESS density matrix in Eqs.~\eqref{eq:ness_cond} and \eqref{eq:stationary_prob_cond} for conditional resetting protocol. The case of unconditional resetting is discussed in the Supplemental Material \cite{SM} and in Ref.~\cite{wald2025stochastic}. Reset states are denoted henceforth as $\ket{+}=\ket{\uparrow_1, \dots \uparrow_N}$ and $\ket{-}=\ket{\downarrow_1, \dots \downarrow_N}$ as in the main text. We assume, without loss of generality, that the system is inizialized in state $\ket{+}$. 

The conditional protocol is defined by the transition matrix $P_{i,j}(n)$ \eqref{eq:P_matrix}, which  determines whether reset state $j=\{+,-\}$ is chosen when a reset is performed after $n\in\mathbb{N}_0$ steps since the last reset to state $\{+,-\}$. For this matrix, we adopt the same protocol of Ref.~\cite{perfetto2021designing}, where the reset state is chosen according to the majority rule: a measurement of the magnetization in the $z$ direction is taken and then one resets the system to either the state $\ket{+}$ or $\ket{-}$ if a majority of spins is found in the up or down direction, respectively. The transition matrix is written according to Born rule for projective measurements as 
\begin{equation}
P_{i,j}(n)=\braket{n_i| \hat{P}_j|n_i}=\braket{i|(U_{\theta}^{\dagger})^n \hat{P}_j (U_{\theta})^n|i}, \quad \mbox{with} \quad i,j \in \{+,-,\} \qquad  \sum_{j=\pm} P_{i,j}(n)=1, \,\, \forall n,  
\label{eq:transition_matrix_conditional}
\end{equation}
with $\ket{n_i}=U^n\ket{i}$. In this equation, $\hat{P}_j$ is the projector onto the subspace of the Hilbert space in which a majority of spins point in the up ($j=+$) or down ($j=-$) Pauli-$z$ direction. Equation \eqref{eq:transition_matrix_conditional} defines a renewal process since the choice of the next reset state $j$ depends only on the previously chosen state $i$, but not on what happened prior to this reset. The many-body density matrix $\rho(t)$ at time $t$ thus obeys the last renewal equation
\begin{align}
\rho(t)&=\sum_{n=0}^t P_{n_+}(t)\ket{n_+}\bra{n_+} + \sum_{n=0}^t P_{n_-}(t)\ket{n_-}\bra{n_-} \nonumber  \\
&= P_{t_+}(t)\ket{t_+}\bra{t_+}+\sum_{m=1}^t P_{(t-m)_+}(t)\ket{(t-m)_+}\bra{(t-m)_+}+\sum_{m=1}^t P_{(t-m)_-}(t)\ket{(t-m)_-}\bra{(t-m)_-},
\label{eq:renewal_eq_two_states}
\end{align}
with $P_{n_{\pm}}(t)$ denoting the probability of staying in state $\ket{n_{\pm}}$ at time $t$. In particular, the term $P_{t_+}(t)$ accounts for trajectories where no reset takes place and the system undergoes unitary evolution for all the $t$ steps. The initial condition gives $P_{t^-}(t)=0$ and the initial resetting probabilities $P_{0_+}(t=0)=1$ and $P_{0_-}(t=0)=0$. The second and third terms, instead, describe cases where the last reset happens at the step $m$ and after that the system undergoes unitary evolution for $t-m$ steps with probability $P_{(t-m)_{\pm}}(t)$ depending on the last reset being to state $\ket{\pm}$, respectively. In order to obtain $\rho(t)$ and its associated long-time limit we thus need to write an expression for $P_{(t-m)_{\pm}}(t)$. This can be done by exploiting the renewal structure of the reset process, which enforces the relation 
\begin{equation}
P_{n_{\pm}}(t)=(1-r_{n-1})P_{(n-1)_{\pm}}(t-1),     \quad n\geq1, t\geq 1.
\end{equation}
An iterative application of the previous equation leads to 
\begin{equation}
P_{n_{\pm}}(t)=\left(\prod_{j=0}^{n-1}(1-r_j) \right) P_{\pm}(t-n)=q_n\, P_{\pm}(t-n), \quad n\geq 1, t\geq 1.
\label{eq:PnP02resets}
\end{equation}
Here we defined the probability $P_{\pm}(t)\equiv P_{0_{\pm}}(t)$ of resetting to state $\ket{\pm}$ at time $t$. This can be written again using the renewal structure of resetting as
\begin{align}
P_{+}(t)&= \sum_{n=0}^{t-1}r_n P_{n_-}(t-1)P_{-,+}(n)+ \sum_{n=0}^{t-1}r_n P_{n_+}(t-1)P_{+,+}(n),  \quad t\geq 1, \label{eq:P0pm}
\end{align}
and analogously for $P_{-}(t)$ by swapping $+$ and $-$ subscripts. The interpretation of Eq.~\eqref{eq:P0pm} is clear. For example, a reset to state $\ket{+}$ happens at time $t$ if at time $t-1$ the system is in the state $\ket{n_{-}}$ and a reset to $\ket{+}$ happens (probability $P_{-,+}(n)$), or if at time $t-1$ the system is in state $\ket{n_+}$ and a reset to state $\ket{0^+}$ takes place (probability $P_{+,+}(n)$). Note that the sum of the two resetting probability is simply $
P_0(t)=P_{0^+}(t)+P_{0^-}(t)$, i.e., the total resetting probability regardless to the chosen reset state. We now insert Eq.~\eqref{eq:PnP02resets} into Eq.~\eqref{eq:P0pm} in order to obtain a closed equation for $P_{+}$ (same equation applies for $P_{-}(t)$ upon swapping the subscripts $+$ and $-$):
\begin{subequations}
\begin{align}
P_{+}(t)=&\sum_{n=1}^{t-1}r_n P_{-,+}(n)P_{-}(t-n-1)q_n+\sum_{n=1}^{t-1}r_n P_{+,+}(n)P_{+}(t-n-1)q_n, \\
&+r_0 P_{-,+}(0)P_{-}(t-1)+ +r_0 P_{+,+}(0)P_{+}(t-1), \quad \mbox{for} \quad t\geq 2. \\
P_{+}(t=1)&=r_0 P_{-,+}(0)P_{-}(0)+r_0 P_{+,+}(0)P_{+}(0)=r_0 P_{+,+}(0).
\label{eq:P0+_time_closed}
\end{align}
\end{subequations}

The convolution structure of the last equation is best simplified by resorting to discrete Laplace transform, which for a generic function $f(t)$ is given by $\hat{f}(z)=\sum_{t=0}^{\infty} z^{-t} f(t)$. For the resetting probabilities $\hat{P}_{\pm}(z)$ in Laplace domain this leads to the following system of linear equations for $\hat{P}_{\pm}(z)$:
\begin{align}
\begin{pmatrix}
\hat{P}_{+}(z) \\ \hat{P}_{-}(z) 
\end{pmatrix}    =\begin{pmatrix}
z r_0 P_{+,+}(0)+z \hat{g}_{+,+}(z) & z r_0 P_{-,+}(0)+z\hat{g}_{-,+}(z) \\
z r_0 P_{+,-}(0)+z \hat{g}_{+,-}(z) & z r_0 P_{-,-}(0)+z\hat{g}_{-,-}(z)
\end{pmatrix}
\begin{pmatrix}
\hat{P}_{+}(z) \\ \hat{P}_{-}(z)
\end{pmatrix}
+\begin{pmatrix}
1 \\
0
\end{pmatrix}.
\label{eq:matrix_eq_P_stat}
\end{align}
Here, we defined the function
\begin{equation}
\hat{g}_{i,j}(z)=\sum_{n=1}^{\infty}z^{-n} R_{i,j}(n)\left(\prod_{j=0}^{n-1}(1-r_j) \right),   
\end{equation}
with $R_{i,j}(n)=r_n P_{i,j}(n)$, as in the main text. Equation \eqref{eq:matrix_eq_P_stat} can simply solved by matrix inversion leading to 
\begin{align}
\begin{pmatrix}
\hat{P}_{+}(z) \\ \hat{P}_{-}(z)
\end{pmatrix}    = \frac{1}{\mbox{det}(\mathbb{I}-A)}\begin{pmatrix}
1-(z r_0 P_{-,-}(0) +z \hat{g}_{-,-}) \\
z r_0 P_{+,-}(0)+z\hat{g}_{+,-},
\end{pmatrix} 
\label{eq:matrix_eq_solved}
\end{align}
with the matrix defined from Eq.~\eqref{eq:matrix_eq_P_stat}
\begin{align}
A &= \begin{pmatrix}
z r_0 P_{+,+}(0)+z \hat{g}_{+,+}(z) & z r_0 P_{-,+}(0)+z\hat{g}_{-,+}(z) \\
z r_0 P_{+,-}(0)+z \hat{g}_{+,-}(z) & z r_0 P_{-,-}(0)+z\hat{g}_{-,-}(z), 
\end{pmatrix},
\label{eq:matrix_det}
\end{align}
From Eqs.~\eqref{eq:matrix_eq_solved} and \eqref{eq:matrix_det}, we are eventually in position of obtained the NESS. To do so, we exploit the final value theorem of Laplace transforms, see, e.g., Ref.~\cite{feller1957introduction}, which relates the stationary limit of the resetting probabilities $P_{\pm}^{\mathrm{stat}}=\lim_{t\to \infty} P_{\pm}(t)$ to the residue associated to the simple pole at $z=1$ in Eq.~\eqref{eq:matrix_det}. The residues associated to $z=1$ are readily computed and yield the stationary values of resetting probability to states $\ket{+}$ (same equation applies for $\ket{-}$ by swapping $+$ and $-$):
\begin{align}
P_{+}^{\mathrm{stat}}=\lim_{z\to 1}(z-1)\hat{P}_{+}(z)=\frac{\hat{g}_{-,+}(1)}{[1+\sum_{n=0}^{\infty}q_{n+1}][r_0 P_{-,+}(0)+r_0 P_{+,-}(0)+\hat{g}_{-,+}(1)+\hat{g}_{+,-}(1) ]},
\label{eq:stationary_rates_conditional}
\end{align}
We remark that the existence of the pole in Eq.~\eqref{eq:stationary_rates_conditional}, and therefore of the stationary state, is not apriori guaranteed. We find that a sufficient condition for the existence of the stationary state is \cite{SM}:
\begin{equation}
\lim_{n \to \infty} n\, q_n=0.
\label{eq:sufficient_condition}
\end{equation}
A decay of the survival probability $q_n$ faster than $1/n$ thus provides a sufficient condition for the existence of the stationary state. As a consistency check, we also note that 
\begin{equation}
P_{+}^{\mathrm{stat}}+P_{-}^{\mathrm{stat}}=P_0^{\mathrm{stat}}=\frac{1}{1+\sum_{n=0}^{\infty}q_{n+1}},   
\end{equation}
which is the stationary probability of performing a reset regardless of the chosen reset state being $\ket{+}$ or $\ket{-}$. This is solely dependent on the waiting time distribution and not on the conditional-measurement protocol, and it, indeed, determines the stationary state in Eq.~\eqref{eq:stationary_state_unc} for the unconditional resetting protocol. The stationary state $\rho^{\mathrm{ness}}$ is then obtained by taking the Z transform of the last renewal equation \eqref{eq:renewal_eq_two_states}. The latter can be written in terms of $\hat{P}_{\pm}(z)$ exploiting Eqs.~\eqref{eq:renewal_eq_two_states} and \eqref{eq:PnP02resets} and the convolution theorem. The long-time limit is again obtained by taking the limit $z\to 1$ as in Eq.~\eqref{eq:stationary_rates_conditional}. This leads to our final expression for the stationary state density matrix in the conditional reset protocol reported in Eq.~\eqref{eq:ness_cond} of the main text.

\clearpage

\def\bra#1{\langle{#1}|}
\def\ket#1{|{#1}\rangle}
\def\braket#1{\langle{#1}\rangle}
\def\Bra#1{\left\langle#1\right|}
\def\Ket#1{\left|#1\right \rangle}
\def\BraVert{\egroup\,\mid\,\bgroup}
\def\lket#1{\vert#1\rangle\hspace{-1mm}\rangle}
\def\lbra#1{\langle\hspace{-1mm}\langle#1\vert}
\def\Brak#1#2#3{\bra{#1}#2\ket{#3}}
\renewcommand{\thefigure}{S\arabic{figure}}
\newcommand{\CB}[1]{{\color{red} CB: #1}}
\setcounter{equation}{0}
\setcounter{figure}{0}
\setcounter{table}{0}
\setcounter{page}{1}
\makeatletter
\renewcommand{\theequation}{S\arabic{equation}}
\renewcommand{\thefigure}{S\arabic{figure}}

\makeatletter
\renewcommand{\theequation}{S\arabic{equation}}
\renewcommand{\thefigure}{S\arabic{figure}}

\renewcommand{\bibnumfmt}[1]{[S#1]}
\renewcommand{\citenumfont}[1]{S#1}

\onecolumngrid

\setcounter{secnumdepth}{3}

\maketitle
\begin{center}
{\Large SUPPLEMENTAL MATERIAL}
\end{center}
\begin{center}
\vspace{0.8cm}
{\Large Nonequilibrium steady states induced by stochastic mid-circuit measurements and resets on a quantum computer}
\end{center}
\begin{center}
Jakob Murauer$^{1}$, Sabine Tornow$^{1}$, and Gabriele Perfetto$^{2}$
\end{center}
\begin{center}
$^1${\em Research Institute CODE, University of the Bundeswehr Munich, 81739 Munich, Germany}\\
$^2${\em Institut für Theoretische Physik, ETH Zürich, Wolfgang-Pauli-Str. 27, 8093 Zürich, Switzerland}\\
\end{center}
\noindent In this Supplemental Material, we present additional details about the theory calculations and the experimental implementations. In Sec.~\ref{sec1:unc_th}, we detail the derivation of the nonequilibrium stationary state density matrix for unconditional resetting. In Sec.~\ref{sec2:floquet_spectrum}, we explain the exact diagonalization of the Floquet transverse-field Ising unitary via mapping to free fermions. We then discuss how degeneracies in the ensuing Floquet spectrum reflects into the stationary-state magnetization for a small number of qubits. In Sec.~\ref{sec3:exp_implementation}, we explain how the unitary gates and stochastic resets are experimentally implemented on the quantum \texttt{ibm\_marrakesh} processor. We discuss both the implementation of the unconditional protocol and the conditional one. In Sec.~\ref{sec4:hardware_calibration}, we report the hardware calibration data relevant to the experiments presented in the main text. In Sec.~\ref{sec5:additional_unc}, we report additional experimental data for unconditional resetting. In Sec.~\ref{sec6:noise_models}, we eventually report a comparison between the experimental data for conditional resetting and various noise models including the noisy reset-state model developed in the main text.

\section{Derivation of the NESS density matrix for unconditional resetting}
\label{sec1:unc_th}
We consider evolution in discrete time according to a unitary gate $U$:
\begin{equation}
\ket{n}=U^n \ket{0}, \quad n \in \mathbb{N}_0=0,1,2 \dots .
\label{eq:unitary_gate}
\end{equation}
All the formulas we write in the following apply for a generic unitary operator. In the previous equation, $\ket{0}$ denotes the reset state, which we take as coinciding with the initial state without loss of generality. We add resetting on the top of the unitary evolution \eqref{eq:unitary_gate}. This means that at each discrete time step $n$ the systems is either reset to the initial state $\ket{0}$, with probability $0\leq r_n \leq 1$ or it undergoes unitary evolution, with probability $1-r_n$. Here, $r_n$ is the equivalent in discrete time of the waiting time distribution for the continuous time implementation of resetting \cite{evans2011resetting,evans2011resetlong,eule2016non,power_law_reset_ness}. The case of poissonian resetting, i.e, resetting at constant rate, corresponds to $r_n=r$, and it has been recently studied in Ref.~\cite{wald2025stochastic}. Here, we extend their analysis by deriving analytical formulas for the nonequilibrium stationary state (NESS) valid for generic waiting time distributions $r_n$. We also derive a sufficient condition that the survival probability $q_n$ must satisfy for the existence of the NESS. 

At discrete time $t$ the many-body state of the system is mixed as a consequence of the resetting process. The associated density matrix $\rho(t)$ is 
\begin{equation}
\rho(t)=\sum_{n=0}^{t} P_n(t)\ket{n}\bra{n}, \quad \mbox{with} \quad \rho(0)=\ket{0}\bra{0},
\label{eq:density_matrix_sm}
\end{equation}
with $P_n(t)$ denoting the probability of staying in state $\ket{n}$ \eqref{eq:unitary_gate} at time $t$. The initial condition for $\rho(0)$ accordingly implies that $P_0(t=0)=1$, and $P_{n\geq 1}(t=0)=0$. We can separate the term $n=t$ in the sum to get
\begin{align}
\rho(t) &= \sum_{n=0}^{t-1}P_n(t)\ket{n}\bra{n} + P_t(t)\ket{t}\bra{t},   \nonumber \\
&= P_t(t)\ket{t}\bra{t} + \sum_{m=1}^{t} P_{t-m}(t)\ket{t-m}\bra{t-m}, \quad n=t-m \quad (\mbox{change of variable in the sum}).
\label{eq:last_ren_discrete}
\end{align}
The last equation is the so-called last renewal equation in the context of discrete time. In particular the term $P_t(t)$ accounts for trajectories where no reset takes place and the system undergoes unitary evolution for all the $t$ steps. The second term, instead, describes cases where the last reset happens at the step $m$ and after that the system undergoes unitary evolution for $t-m$ steps with probability $P_{t-m}(t)$. The long-time limit of $P_{n}(t)$ can be derived by relating this quantity to the probability $P_0(t)$ of performing a reset at time $t$ (without knowledge of when the previous reset took place). The relation between $P_n(t)$ and $P_0(t)$ is dictated by the renewal structure of the reset process. Namely, one has
\begin{equation}
P_n(t)=(1-r_{n-1})P_{n-1}(t-1), \quad n\geq 1, t\geq 1, 
\label{eq:ren_1state_n}
\end{equation}
which states that the state $\ket{n}$ is populated at time $t$ if the state $\ket{n-1}$ was populated at time $t-1$ and no reset event takes place afterwards (with probability $1-r_{n-1}$). The previous equation can be iteratively applied to get  
\begin{equation}
P_n(t)=\left(\prod_{j=0}^{n-1} (1-r_j)\right) P_0(t-n)=q_n P_0(t-n),
\label{eq:ren_1state_n2}
\end{equation}
where we defined the survival probability $q_n$ 
\begin{equation}
q_n=\prod_{j=0}^{n-1}(1-r_j),    
\label{eq:survival_probability}
\end{equation}
which gives the probability that the next reset takes place after $n$ steps from the previous reset. The asymptotic decay of $q_n$ as a function of $n$ is important in order for the NESS to exist, as we now establish. The probability $P_n(t)$ is fully determined once the probability $P_0(t)$ is known. This can be written in terms of $P_n(t)$ as 
\begin{equation}
P_0(t) = \sum_{n=0}^{t-1} r_n P_{n}(t-1), \quad t\geq 1.
\label{eq:ren_1state_0}
\end{equation}
This equation states that a reset takes place at the step $t$ after the system reached the state $\ket{n-1}$ at the previous step $t-1$. Note that the previous equation applies for $t\geq 1$, while for $t=0$ we have the initial condition $P_0(t=0)=1$ reported above. Inserting Eq.~\eqref{eq:ren_1state_n2} into \eqref{eq:ren_1state_0} we get a closed equation for $P_0(t)$:
\begin{align}
P_0(t)&= \sum_{n=1}^{t-1} r_n q_n \, P_0(t-n-1) + r_0 P_0(t-1), \quad \mbox{for} \quad t\geq 2, 
\label{eq:ren_0_convolution}
\\
P_0(t)&=r_0 P_0(t-1), \quad \mbox{for} \quad t=1. 
\end{align}

Equation Eq.~\eqref{eq:ren_0_convolution} is in the convolution form. It is therefore convenient to simplify this using the discrete Laplace transform, also known as unilateral $Z$ transform, see, e.g., Ref.~\cite{feller1957introduction}. In particular, we have
\begin{equation}
\hat{P}_0(z)=\sum_{t=0}^{\infty}z^{-t} P_0(t)= P_0(0)+z^{-1}P_0(1)+\sum_{t=2}^{\infty}z^{-t}P_0(t),
\label{eq:gen_Z_P0}
\end{equation}
In the last sum in Eq.~\eqref{eq:gen_Z_P0} we use \eqref{eq:ren_0_convolution}:
\begin{align}
\sum_{t=2}^{\infty}z^{-t}r_0 P_0(t-1)&= r_0 \sum_{t=1}^{\infty}z^{-t}z^{-1}P_0(m)=r_0 z^{-1}(\hat{P}_0(z)-1),  \\
\sum_{t=2}^{\infty} z^{-t} \sum_{n=1}^{t-1} r_n q_n P_0(t-n-1)&= z^{-1}\hat{P}_0(z)\hat{g}(z),
\label{eq:intermediate_Z_1reset}
\end{align}
In the previous equation, we defined the $Z$ transform $\hat{g}(z)$
\begin{equation}
\hat{g}(z)= \sum_{n=1}^{\infty}r_n q_n \, z^{-n},  \, \, \mbox{for} \, \, n \geq 1.
\label{eq:g_fun_z}
\end{equation}
while $g_{n=0}\equiv 0$. The factor $z^{-1}$ in \eqref{eq:intermediate_Z_1reset} comes from the shifting theorem of $Z$ transform since the sum runs up to $t-1$. We insert Eq.~\eqref{eq:intermediate_Z_1reset} into Eq.~\eqref{eq:gen_Z_P0} and we get a closed equation for $\hat{P}_0(z)$
\begin{align}
\hat{P}_0(z)&=1 +r_0 z^{-1} + r_0 z^{-1} \hat{P}_0(z)-r_0 z^{-1} +z^{-1}\hat{P}_0(z)\hat{g}(z)=1+z^{-1}\hat{P}_0(z)(r_0+\hat{g}(z)),\nonumber \\
\hat{P}_0(z)&=\frac{z}{z-(r_0+\hat{g}(z))}.
\label{eq:gen_Z_final_1reset}
\end{align}
The last equation crucially allows to obtain the stationary limit of $P_0(t)$ provided it exists:
\begin{equation}
P_0^{\mathrm{stat}}=\lim_{t \to \infty} P_0(t).
\label{eq:stationary_resetting_probability}
\end{equation}
The latter limit, in turn, fully identifies the stationary state density matrix $\rho^{\mathrm{ness}}$, which can be obtained by taking the $Z$ transform of Eq.~\eqref{eq:last_ren_discrete} using Eq.~\eqref{eq:ren_1state_n2} and the convolution theorem:
\begin{equation}
\hat{\rho}(z)=\sum_{t=0}^{\infty} z^{-t} P_t(t)\ket{t}\braket{t}+   \hat{P}_0(z)\sum_{t=1}^{\infty} z^{-t} q_t\, \ket{t}  \braket{t}+\hat{P}_0(z)\ket{0}\bra{0}.
\label{eq:Z_transform_renewal_eq}
\end{equation}
We then use the final value theorem of $Z$ transforms, which gives the long time limit of $\rho(t)$ in terms of the singular behavior of $\hat{\rho}(z)$ at $z=1$ \cite{feller1957introduction}. It amounts to determine whether $\hat{\rho}(z)$ and $\hat{P}_0(z)$ have a simple pole at $z=1$ or not. From this we get
\begin{equation}
\rho^{\mathrm{ness}}=\lim_{t\to\infty}\rho(t)=\lim_{z\to 1} (z-1)\hat{\rho}(z) = P_0^{\mathrm{stat}} \sum_{t=1}^{\infty}q_t \ket{t}\bra{t} + P_0^{\mathrm{stat}}\ket{0}\bra{0}. 
\label{eq:stationary_state_1}
\end{equation}
This equation coincides with Eq.~(2) of the main text.

Here we assumed that the first on the right hand side of \eqref{eq:Z_transform_renewal_eq} has no pole in $z=1$, while $P_0^{\mathrm{stat}}$ is the residue at $z=1$ associated to the pole of \eqref{eq:gen_Z_final_1reset}. For a pole in $z=1$ to exist, we need a zero in the denominator of $\hat{P}_0$ and therefore the following equation must hold true
\begin{equation}
r_0+\hat{g}(z=1)=1.
\label{eq:stationary_state_condition}
\end{equation}
We now see which conditions Eq.~\eqref{eq:stationary_state_condition} enforces on the waiting time distribution $r_n$ in order for the stationary limit \eqref{eq:stationary_resetting_probability} and \eqref{eq:stationary_state_1} to exist. From the definition in Eq.~\eqref{eq:g_fun_z} evaluated at $z=1$ we have
\begin{align}
\hat{g}(z=1) &=\sum_{n=1}^{\infty}r_n q_n= \sum_{n=1}^{\infty}r_n \left( \prod_{j=0}^{n-1} (1-r_j) \right)=\sum_{n=1}^{\infty} q_n -\sum_{n=1}^{\infty}  q_{n+1} \nonumber\\
&= q_1 -\lim_{n\to \infty} q_{n+1}=1-r_0 -\lim_{n\to \infty}  q_{n+1}.
\label{eq:pole_condition_survival}
\end{align}
In order for Eq.~\eqref{eq:stationary_state_condition} to be satisfied we necessarily need that the survival probability 
\begin{equation}
\lim_{n\to \infty}q_n=0,
\label{eq:condition_1}
\end{equation}
decays to zero at large $n$. From the physics point of view, this condition implies that reset realizations where consecutive reset events are separated by a large sequence of unitary steps are unlikely to happen. This is intuitively clear: reset events are necessary to reach a stationary state since unitary dynamics alone cannot drive the system to any stationary state. Equation \eqref{eq:condition_1} is, however, necessary but not sufficient for the existence of the stationary state. The time interval between consecutive resets must, indeed, be small enough so that a large number of resets fits into a typical trajectory and stationary is eventually attained. This requires the survival probability to decay sufficiently fast. To quantify this aspect, we can eventually compute the residue of $\hat{P}_0(z)$ at $z=1$:
\begin{equation}
P_0^{\mathrm{stat}}=\lim_{z-1}(z-1)\hat{P}_0(z)= \frac{1}{1-\hat{g}'(z=1)},    
\label{eq:residue_1}
\end{equation}
and 
\begin{align}
-\hat{g}'(z=1)&=\sum_{n=1}^{\infty}n r_n q_n=\sum_{n=1}^{\infty}n  (q_n-q_{n+1})=\sum_{n=1}^{\infty}n q_n -\sum_{n=1}^{\infty}(n+1)q_{n+1} + \sum_{n=1}^{\infty}q_{n+1}, \\
&=q_1-\lim_{n\to \infty}(n+1)q_{n+1}+\sum_{n=1}^{\infty}q_{n+1}=1-r_0 + \sum_{n=1}^{\infty} q_{n+1}.
\label{eq:survival_sum}
\end{align}
A sufficient condition for the residue in Eq.~\eqref{eq:residue_1} to be finite is thus given by 
\begin{equation}
\lim_{n\to \infty}n\, q_n =0.
\label{eq:condition_2}
\end{equation}
We establish Eq.~\eqref{eq:condition_2} as a sufficient condition such that stationary is reached and Eqs.~\eqref{eq:stationary_resetting_probability} and \eqref{eq:stationary_state_1} apply. It implies that $q_n$ decays asymptotically at large $n$ faster than $1/n$, which ensures that the sum in Eq.~\eqref{eq:survival_sum} is convergent. Equation \eqref{eq:condition_2} also ensures that $\hat{q}(z)$ is analytic in $z=1$, as we assumed in the derivation of the stationary state \eqref{eq:stationary_state_1}. We note that a condition analogous to Eq.~\eqref{eq:condition_2} for the existence of the stationary state has been derived for a classical diffusion process subject to resetting in continuous time in Ref.~\cite{power_law_reset_ness}. Therein, a stationary state is similarly found only if the continuous-time waiting time distribution $p(\tau)$ (survival probability $q(\tau)$) decays faster than $1/\tau^2$ ($1/\tau$) for large time $\tau$ between consecutive resets. Assuming the validity of the conditions \eqref{eq:condition_1} and \eqref{eq:condition_2} we eventually get in Eq.~\eqref{eq:residue_1}
\begin{equation}
P_0^{\mathrm{stat}}= \frac{1}{1+\sum_{n=0}^{\infty}\prod_{j=0}^{n}(1-r_j)}=\frac{1}{1+\sum_{n=0}^{\infty}q_{n+1}}.
\label{eq:P_0_stat}
\end{equation}
Equations \eqref{eq:stationary_state_1}, \eqref{eq:condition_2} and \eqref{eq:P_0_stat} give the exact expression of the steady state for the unitary dynamics in discrete time in the presence of resetting with arbitrary waiting time distribution $r_n$.

As a final benchmark, we can specialize the previous formulas to poissonian resetting, where the waiting time distribution is constant $r_n=r$. In that case the survival probability \eqref{eq:survival_probability} reads as 
\begin{equation}
q_n=\prod_{j=0}^{n-1}(1-r_j)=(1-r)^n.
\label{eq:poissonian_survival}
\end{equation}
It decays exponentially to zero as a function of $n$ and therefore both conditions \eqref{eq:condition_1} and \eqref{eq:condition_2} are met. Poissonian resetting thus drives the dynamics towards a NESS. The stationary resetting probability \eqref{eq:stationary_resetting_probability} is
\begin{equation}
P_0^{\mathrm{stat}}= \frac{1}{1+\sum_{n=0}^{\infty}q_{n+1}} =\frac{1}{1+\frac{1-r}{1-(1-r)}}=r.
\label{eq:poissonian_reset_probability}
\end{equation}
This identity simply reflects the markovian nature of poissonian resetting, where the probability of performing a reset at a given time $t$ is time independent $P_0(t)=r$ regardless of the time elapsed since the previous reset. Plugging \eqref{eq:poissonian_survival} and \eqref{eq:poissonian_reset_probability} into Eq.~\eqref{eq:stationary_state_1} one has
\begin{equation}
\rho^{\mathrm{ness}}=r\sum_{t=1}^{\infty}(1-r)^t \ket{t}\bra{t}+r\ket{0}\bra{0}=r\sum_{t=0}^{\infty}(1-r)^t\ket{t}\bra{t}.
\label{eq:stationary_state_poissonians}
\end{equation}
This is in agreement with the expression of Ref.~\cite{wald2025stochastic}, which is here derived as a particular case of our general results. Equation \eqref{eq:stationary_state_poissonians} has been used in the main text to derive the noiseless predictions for the unconditional resetting protocol (solid blue lines in Fig.~2 of the main text).

\section{Floquet transverse-field Ising model}
\label{sec2:floquet_spectrum}
In Subsec.~\ref{subsec2.1:spectrum}, we detail the derivation of the exact Floquet spectrum of the unitary in Eqs.~(6) and (7) of the main text. In Subsec.~\ref{subsec2.2:NESS_degeneracies}, we discuss the effect of such degeneracies on the NESS expectation value of the magnetization in the unconditional resetting protocol for $N=3$ qubits.

\subsection{Exact spectrum of the Floquet transverse-field Ising model}
\label{subsec2.1:spectrum}
The derivation of the eigenspectrum of the Floquet transverse field Ising model via Jordan-Wigner transformation and mapping to free fermions is a standard technique, see, e.g., Refs.~\cite{FTIM_1,FTIM_2,FTIM_3}. We therefore briefly review here only the main aspects of the derivation that are necessary to understand the NESS obtained in the presence of resetting for $N=3$ qubits.

The Ising unitary gates are given by (cf. Eqs.~(6) and (7) of main text) by
\begin{equation}
U_{\theta}= e^{-i H_x \theta}  e^{-i H_{zz}\theta}=  U_x U_{zz}, \quad \mbox{with} \quad U_{x}=e^{-i H_x \theta}, \mbox{and} \quad U_{zz}=e^{-i H_{zz}\theta}.
\label{eq:ising_discrete_unitary_sm}
\end{equation}
with 
\begin{equation}
H_x=-J h\sum_{i=1}^{N}\sigma^x_i, \quad \mbox{and} \quad H_{zz}=-J\sum_{i=1}^{N}\sigma^z_i \sigma^z_{i+1}.
\label{eq:ising_discrete_terms_sm}
\end{equation}
We map these gates to fermions exploiting the Jordan-Wigner transformation
\begin{equation}
\sigma^x_j=2 c_j^{\dagger}c_j-1, \quad \sigma_j^z=-\left(\prod_{l<j} (1-2c_l^{\dagger}c_l) \right) (c_j+c_j^{\dagger}),
\label{eq:JW_transformation}
\end{equation}
where $c_j$ ($c_j^{\dagger}$) are fermionic annihilation (creation) operators at site $j$ and satisfy canonical anticommutation relation $\{c_j,c_{j'}^{\dagger}\}=\delta_{j,j'}$. The spin operators obey periodic boundary conditions (PBC) $\sigma_{j+N}^{\alpha}=\sigma^{\alpha}_j$. We insert Eq.~\eqref{eq:JW_transformation} into Eqs.~(6) and (7) and we get
\begin{equation}
U_{zz}=\mbox{exp}\left(i\theta J\sum_{j=1}^N (c_j^{\dagger}c_{j+1}^{\dagger}-c_j c_{j+1}+c_j^{\dagger}c_{j+1}-c_jc_{j+1}^{\dagger})\right), \quad U_x=\mbox{exp}\left(i\theta J h\sum_{j=1}^N (2c_j^{\dagger}c_j -1) \right).
\label{eq:Ising_gates_real_space}
\end{equation}
In the gate $U_{zz}$ in the previous equation it is important to specify the boundary conditions obeyed by the fermionic operators. These depend on the parity $P=(-1)^{\sum_{j=1}^N c_j^{\dagger}c_j}$ of the total number of fermions : $c_{j+N}=c_j$ (PBC) if $P=-1$ and $c_{j+N}=-c_j$ (APBC) if $P=1$. The boundary conditions for the fermions, in turn, determine the allowed values of the momentum $q$ when Fourier transform is taken:
\begin{equation}
c_j=\frac{e^{-i\pi/4}}{\sqrt{N}}\sum_{n} e^{i q_n j} c_q,
\label{eq:Fourier_transform}
\end{equation}
with $q_n=2\pi n/N$ for $P=-1$ and $q_n=2\pi (n+1/2)/N$ for $P=1$. The variable $n$ runs over the set of integers $n\in \{-[N/2], -[N/2]+1,\dots, [N/2]-1, [N/2] \}$, where we consider an odd number $N$ of qubits as in the main text and $[N/2]$ denotes the integer part. In this case, the odd sector $P=-1$ contains the mode $q=0$, while the even sector $P=+1$ the mode $q=\pi$. The modes $q=0$ and $\pi$ must be treated separately in the diagonalization of the unitary since they do not have a negative counterpart in the Brillouin zone of the allowed momenta ranging in $(-\pi,\pi]$. Using Eq.~\eqref{eq:Fourier_transform} into Eq.~\eqref{eq:Ising_gates_real_space} we can write the Ising unitary gates as quadratic forms of the fermionic spinors $\phi_q=(c_q,c^{\dagger}_{-q})$ [$\phi_q^{\dagger}=(c_q^{\dagger},c_{-q})$] in momentum space
\begin{align}
U_{\theta}&= U_{>0} U_{\mathrm{0}}, \quad U_{>0}=\prod_{q>0}\mbox{exp}(-i\phi_q^{\dagger}X_q \phi_q) \, \mbox{exp}(-i \phi_q^{\dagger} Z_q \phi_q), \label{eq:Ising_gates_momentum_space_bulk} \\
U_0&=\mbox{exp}[i\theta J (2 c_0^{\dagger}c_0 -1)(1-P)/2-i\theta J (2 c_{\pi}^{\dagger}c_{\pi} -1)(1+P)/2
+i\theta Jh (2 c_0^{\dagger}c_0(1-P)/2 + 2 c_{\pi}^{\dagger}c_{\pi}(1+P)/2 -1)],
\label{eq:Ising_gates_momentum_space_boundary}
\end{align}
with the matrices $X_q$ and $Z_q$ given by 
\begin{equation}
Z_q=-2J\theta \begin{pmatrix}
\cos q & i\sin q \\ 
-i \sin q & -\cos q
\end{pmatrix}, \quad X_q= -2J h\theta 
\begin{pmatrix}
 1 & 0 \\
 0 & -1
\end{pmatrix}.
\end{equation}
In Eq.~\eqref{eq:Ising_gates_momentum_space_bulk}, we factorized in $U_{>0}$ different positive momenta $q>0$ since they commute between each other. The momenta $q=0$ and $q=\pi$ are treated separately in the unitary $U_0$ in Eq.~\eqref{eq:Ising_gates_momentum_space_boundary}. The term $U_0$ is already diagonal in the fermionic operators and commutes with $U_{>}$. The latter is not diagonal since pairs of opposite modes are therein coupled. In order to bring $U_{>}$ to a diagonal form, a Bogoliubov rotation is eventually performed. This amounts to determining a unitary matrix $S$ such that
\begin{equation}
S [\mbox{exp}(-i X_q) \mbox{exp}(-i Z_q)]S^{-1}=\begin{pmatrix}
\mbox{exp}(-i \varepsilon_q) & 0 \\ 0 & \mbox{exp}(i \varepsilon_q) 
\end{pmatrix},  \quad \phi_q=\begin{pmatrix} c_q \\ c^{\dagger}_{-q}
\end{pmatrix}=S \begin{pmatrix}
    d_q \\ d_{-q}^{\dagger}
\end{pmatrix}=\begin{pmatrix}
 u_q & v_q \\ -v_q^{\ast} & u_q^{\ast}   
\end{pmatrix} \begin{pmatrix}
    d_q \\ d_{-q}^{\dagger}
\end{pmatrix},
\label{eq:bogoliubov_1}
\end{equation}
where $d_q$ ($d_q^{\dagger}$) are the Bogoliubov annihilation (creation) operators for mode $q$ and satisfy canonical anticommutation relations $\{d_q, d_{q'}^{\dagger}\}=\delta_{q,q'}$. For the modes $q=0$ and $q=\pi$ no Bogoliubov rotation is necessary and therefore $c_{0,\pi}=d_{0,\pi}$ and $c_{0,\pi}^{\dagger}=d_{0,\pi}^{\dagger}$. The coefficients $u_q$ and $v_q$ of the unitary transformation are found by direct calculation and read as 
\begin{align}
u_q&=\frac{b_q}{\sqrt{2\xi_q(\xi_q+a_q)}}, \quad v_q= \frac{b_q}{\sqrt{2\xi_q(\xi_q-a_q)}}, \\
a_q&= \cos(2 J\theta)\sin(2Jh\theta)+\cos(2Jh\theta)\sin(2J\theta)\cos q, \quad b_q=-\sin q \sin(2J\theta) e^{i2 Jh\theta}, \quad  \xi_q =a_q^2+|b_q|^2.
\label{eq:bogoliubov_2}
\end{align}
The unitary $U_{>}$ eventually appears in a diagonal form in terms of the Bogoliubov fermions $d_q$, with the Floquet Hamiltonian $H_F$  and its eigenspectrum $\varepsilon_q$ given by
\begin{align}
U_{>0}&=\mbox{exp}(-i H_F), \quad H_F=\sum_{q}\varepsilon(q)\left(d_q^{\dagger}d_q-\frac{1}{2}\right), \\
\varepsilon(q) &=|\arccos(\cos(2J\theta)\cos(2 Jh\theta)-\sin(2J\theta) \sin(2h\theta)\cos q )|.
\label{eq:bogoliubov_3}
\end{align}
We note that in Eq.~\eqref{eq:bogoliubov_1} we can consider the diagonalization of each $q$ block as matrix, and not as an operator, since the product of two gaussians in the fermionic bispinors $\phi_q$ is still a gaussian in the binospors
\begin{equation}
\mbox{exp}(-i\phi_q^{\dagger}X_q \phi_q) \, \mbox{exp}(-i \phi_q^{\dagger} Z_q \phi_q)=\mbox{exp}(-i \phi_q^{\dagger} A_q \phi_q), \quad \mbox{with} \quad \mbox{exp}(-i X_q) \, \mbox{exp}(-i  Z_q)=\mbox{exp}(-i A_q),
\label{eq:gaussian_identity}
\end{equation}
as can be proved using the Baker–Campbell–Hausdorff formula \cite{FTIM_1,FTIM_2,FTIM_3}. From the Floquet dispertion relation in Eq.~\eqref{eq:bogoliubov_3}, one can also see that the mode $q=\pi$ becomes gapless when $h=1$. This reflects the quantum phase transition taking place in the transverse-field Ising model (whose Hamiltonian $H_x+H_{zz}$ is obtained in the limit of small $\theta$) at zero temperature in the thermodynamic limit. 

\subsection{Effect of the degeneracies in the spectrum for the magnetization NESS of few qubits}
\label{subsec2.2:NESS_degeneracies}
In Fig.~2(a), we observe that for $N=3$ qubits the stationary magnetization features two local maxima at $h\approx 0.56$ and $h\approx 1.01$. In this section, we link the appearance of these local maxima in $h$ to degeneracies in the exact spectrum in Eq.~\eqref{eq:bogoliubov_3}.

To do this, we decompose the NESS expectation of a generic operator $\mathcal{O}$ for unconditional resetting (a similar analysis can be done for conditional resetting) according to Eq.~\eqref{eq:stationary_state_poissonians} in terms of the eigenstates $\ket{E_{\theta}}$ of the unitary $U_{\theta}$: $U_{\theta}\ket{E_{\theta}}=\mbox{exp}(-i E_{\theta})\ket{E_{\theta}}$. Here $E_{\theta}$ are the so-called Floquet quasi-energies that can be determined from the diagonalized unitary in Eq.~\eqref{eq:bogoliubov_3} and Eq.~\eqref{eq:Ising_gates_momentum_space_boundary} for the boundary term $U_0$. We then take the expectation value of the observable $\mathcal{O}$ over the state \eqref{eq:stationary_state_poissonians} and we insert twice the resolution of the identity in terms of the Floquet eigenstates to get 
\begin{equation}
\braket{\mathcal{O}}_{\mathrm{ness}}^{\mathrm{uncond.}}=\mbox{Tr}[\mathcal{O}\rho_{\mathrm{ness}}]=r\sum_{t=0}^{\infty}(1-r)^t \Braket{0|(U_{\theta}^n)^{\dagger}\, \mathcal{O} \, U_{\theta}^n |0}=\sum_{E_{\theta},E_{\theta}'}f_r(E_{\theta},E_{\theta}')\braket{0|E_{\theta}}\braket{E_{\theta}'|0}\mathcal{O}_{E_{\theta},E_{\theta}'}.  
\label{eq:ness_spectral_overlap_0}
\end{equation}
Here, we defined the function $f_r(E_{\theta},E_{\theta}')$ which depends on the quasi-energies $E_{\theta},E_{\theta}'$ of the eigenstates and the overlap $\mathcal{O}_{E_{\theta},E_{\theta}'}$ of the observable of interest with respect to such eigenstates:
\begin{equation}
f_r(E_{\theta},E_{\theta}')=\frac{r}{1-(1-r)e^{i(E_{\theta}-E_{\theta}')}}, \quad \mathcal{O}_{E_{\theta},E_{\theta}'}=\braket{E_{\theta}|\mathcal{O}|E_{\theta}'}.
\label{eq:ness_spectral_overlaps}
\end{equation}
Importantly, we observe that $|f_r(E_{\theta},E'_{\theta})|\leq 1$, with equality achieved only when $E_{\theta}-E_{\theta}'=2\pi p$, with $p$ an integer number. Henceforth, we define two Floquet quasi-energies to be resonant when such condition is met, since Floquet quasi-energies are defined only modulo $2\pi$. Equation \eqref{eq:ness_spectral_overlaps} implies that a resonance of two, or more, quasienergies is a necessary condition for a maximum in the NESS expectation value to appear. This condition is, however, not sufficient for a maximum in the NESS expectation value to develop, since also the overlaps between the reset state and the eigenstates, and the matrix element of the operator carry an energy dependence in Eq.~\eqref{eq:ness_spectral_overlaps}. 

In the case of the Floquet transverse-field Ising model, we can check for the occurrence of resonances in the spectrum directly from the exact knowledge of the Floquet Hamiltonian in Eq.~\eqref{eq:Ising_gates_momentum_space_boundary} and \eqref{eq:bogoliubov_3}. The Floquet eigenenergies $E_n^{\mu}$ for $N=3$ qubits are labelled by the number $n=0,1,2,3$ of fermions in the state and by the index $\mu$, which accounts for different states with the same number $n$ of fermions. The parity $(-1)^n$ of the fermions is also important since it determines the allowed wavevectors as discussed after Eq.~\eqref{eq:Fourier_transform}. With this information at our disposal, we can eventually write down the analytical expression of all the Floquet eigenenergies: 
\begin{subequations}
\begin{align}
E_0&=-\varepsilon(q=\pi/3)+J(h-1)\theta, \\
E_2^{1}&=\varepsilon(q=\pi/3)+J(h-1)\theta, \\
E_2^{2}&=E_2^{3}=J(1-h)\theta,\\
E_3&=\varepsilon(q=2\pi/3)-J(h+1)\theta \\
E_1^1&=E_1^2=J(1+h)\theta,\\
E_1^3&=-\varepsilon(q=2\pi/3)-J(h+1)\theta.
\end{align}
\label{eq:spectrum_N_3}%
\end{subequations}
By analyzing these eigenvalues as a function of the transverse field $h$, we observe that at the peak $h\approx 0.56$ there is a crossing of the eigenvalues $E_0=E_3$, while for $h\approx 1.01$ the eigenvalues $E_2^1=E_1^3$ cross. All other eigenvalues are, instead, distinct at these two $h$ values (apart from $E_2^2=E_2^3$ and $E_1^1=E_1^2$ regardless of the value of $h$). This result establishes a direct link between the local maxima observed in the NESS of the magnetization density and the degeneracies in the Floquet eigenspectrum. The latter, of course, becomes dense in the thermodynamic limit $N\to \infty$ of an infinite number of qubits and the effect of additional degeneracies tends to fade out as $N$ increases, in agreement with Fig.~2(b) and (c). It is important to note that the Floquet spectrum \eqref{eq:spectrum_N_3} exhibits numerous eigenphases crossings, not only those mentioned above for $h\approx 0.56$ and $h\approx 1.01$. As mentioned above, only a select subset of eigenvalue crossings, indeed, manifests in the NESS of the order parameter $\braket{m}_{\mathrm{ness}}^{\mathrm{unc.}}$. Whether a specific crossing induces a maximum in the NESS is also determined by the matrix elements of the observable of interest and by the overlap of the initial state with the Floquet eigenstates, according to Eq.~\eqref{eq:ness_spectral_overlap_0} and \eqref{eq:ness_spectral_overlaps}. In other words, different physically relevant observables, e.g., magnetization correlation functions $\braket{\sigma^z_j \sigma^z_{j+r}}-\braket{\sigma^z_j}\braket{\sigma^z_{j+r}}$, are expected to be sensible to resonances occurring at different $h$ values. We leave the investigation of such features to future work.

\section{Details on quantum hardware experiments}
\label{sec3:exp_implementation}

To simulate the reset protocol on superconducting hardware, we evaluate an ensemble of independent quantum trajectories. Crucially, we construct and execute each stochastic trajectory as an independent quantum circuit. In Subsec.~\ref{sec:Unitary implementation}, we detail the implementation of the unitary time evolution given by the Floquet transverse-field Ising model. In Subsecs.~\ref{sec:unconditional_impl} and \ref{sec:cond_impl} we explain the implementation of unconditional and conditional resetting, respectively. We also explain the error mitigation techniques exploited for unconditional resetting and the implementation of the measurement protocol for conditional resetting.

\subsection{Unitary implementation}
\label{sec:Unitary implementation}
By default, the quantum processor natively initializes each trajectory circuit with all $N$ qubits prepared in the computational zero state $\ket{0}^{\otimes N} = \ket{\uparrow}^{\otimes N}$. We adopt henceforth the notation used in quantum computation \cite{nielsen2010quantum} where $\ket{0}=\ket{\uparrow}=(1,0)$ (shorthand for a column vector) and $\ket{1}=\ket{\downarrow}=(0,1)$ (shorthand for a column vector), where $\ket{\uparrow}$ and $\ket{\downarrow}$ are positive and negative eigenstates of the matrix $\sigma^z$ of the Pauli basis. The unitary dynamics is given by the Floquet transverse-field Ising model in Eqs.~\eqref{eq:ising_discrete_unitary_sm} and \eqref{eq:ising_discrete_terms_sm}, where we assume use periodic boundary conditions (PBC) $\sigma^z_{i+N}=\sigma^z_i$. The last spin of the chain $\sigma^z_N$ is therefore coupled to the first one $\sigma^z_1$.

\begin{figure}[H]
    \centering
    \includegraphics[width=0.5\textwidth]{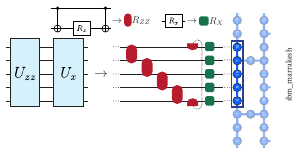}
    \caption{\textit{Quantum circuit implementation of the unitary operations on the quantum computer.} The diagram details the decomposition of two-qubit $U_{zz}$
  interactions into into native $\mathrm{CNOT}$, $\mathrm{RZ}$ gates. This yields a $R_{ZZ}$ gate (red block). Single-qubit $U_{x}$	
  operations are implemented via native $\mathrm{RX}$ gates (green blocks). The mapping on the physical qubits on the hardware machine is reported on the right for $N=5$ qubits.}
    \label{fig:unitary_gates_impl}
\end{figure}

Unitary execution consists of a layer of two-qubit $U_{zz}$ interactions followed by 
a layer of single-qubit $U_x$ operations. The latter is straightforward to implement 
via the native $\mathrm{RX}$ gates available on the superconducting device as 
$U_{x}=\exp(-i\theta \sigma^x_i)=\mathrm{RX}_i(2\theta)$. For the two-qubit 
interactions, although the hardware supports native $\mathrm{R_{ZZ}}$ gates, our 
preliminary hardware benchmarking indicated that the native gates suffer from higher 
cross-talk and lower coherent control fidelity. Consequently, decomposing the 
interaction into standard controlled-NOT ($\mathrm{CNOT}$) gates --- defined as 
$\mathrm{CNOT} = |0\rangle\langle0| \otimes I + |1\rangle\langle1| \otimes \sigma^x$ --- and 
single-qubit rotations $\mathrm{RZ}_i(2\theta)=\exp(-i\theta\sigma^z_i)$ yields 
more reliable results in preliminary reset protocol experiments on hardware. We therefore implement the two-qubit interaction gate $U_{zz}$ as
\begin{align}
    U_{zz}=e^{-i\theta \sigma^z_i \sigma^z_{i+1}} = \mathrm{CNOT}_{i\rightarrow i+1} \mathrm{RZ}_{i+1}(2\theta) \mathrm{CNOT}_{i\rightarrow i+1}.
\end{align}
Because our model dictates periodic boundary conditions (PBC) to form a 1D ring, the limited native connectivity of the heavy-hex hardware lattice is insufficient to directly couple the ends of the chain. Consequently, we introduce $\mathrm{SWAP}$ gates to route the necessary non-local interactions. The overall gate decomposition, optimal qubit mapping, and minimization of the total circuit depth are handled by Qiskit's transpiler optimization routines. In Fig.~\ref{fig:unitary_gates_impl}, we report the quantum circuit implementation of the Ising gates together with the physical depiction of the qubits on the hardware processor.

\subsection{Unconditional resets}
\label{sec:unconditional_impl}
For the unconditional reset protocol, we circumvent active hardware resets by simulating the system dynamics through pure unitary evolution corresponding to the time elapsed since the last stochastically assigned reset event. Our primary objective is to estimate the expectation value of the magnetization $\langle m \rangle = \sum_i \langle \sigma_i^z \rangle/N$. Since unconditional resetting requires no active mid-circuit interventions, we can extensively apply expectation-value-based quantum error mitigation (QEM) techniques. Specifically, we utilize dynamical decoupling (DD), readout error mitigation via twirled readout error extinction (TREX), zero-noise extrapolation (ZNE), and Pauli twirling to suppress hardware noise without disrupting the simulated trajectories. We briefly detail the theoretical framework for each of these techniques here below:
\begin{itemize}
    \item \textbf{Dynamical Decoupling (DD):} To suppress low-frequency environmental noise and cross-talk 
during idle periods, we apply periodic sequences of $\pi$-pulses (for instance Pauli $X$ 
gates)~\cite{DD1, DD2}. One interleaves the natural system-environment evolution 
$U_{\mathrm{SE}}(\tau) = \exp(-iH_{\mathrm{SE}}\tau)$ with pulse operators $P_k$, which 
are elements of the decoupling group chosen to average out specific noise terms (e.g., 
$P_k \in \{I, \sigma^x, \sigma^y, \sigma^z\}$). The resulting effective evolution over a cycle 
$T_c$ averages the system-environment coupling to zero:
\begin{equation}
U_{\mathrm{eff}}(T_c) = \prod_{k} P_k U_{\mathrm{SE}}(\tau_k) P_k^\dagger \approx I, \tag{S47}
\end{equation}
thereby dynamically decoupling the target qubits from the bath. 
    \item \textbf{Pauli Twirling:} To mitigate the impact of coherent over-rotations and structured cross-talk, we employ Pauli twirling \cite{PhysRevLett.82.2417DD, PhysRevLett.119.180509ZNE, PhysRevA.94.052325twirling, PhysRevA.105.032620TREX} to transform these adversarial effects into a well-behaved stochastic Pauli channel. Let $\mathcal{E}$ denote the completely positive trace-preserving (CPTP) map describing the noise associated with the execution of a target unitary, acting on an $n$-qubit density matrix $\rho$. By conjugating this error channel with uniformly sampled operators $P$ from the $n$-qubit Pauli group $\mathcal{P}$, the effective channel becomes:
    \begin{equation}
        \mathcal{E}_{\text{twirled}}(\rho) = \frac{1}{|\mathcal{P}|} \sum_{P \in \mathcal{P}} P^\dagger \mathcal{E}(P \rho P^\dagger) P.
    \end{equation}
This procedure effectively removes the off-diagonal elements of the process matrix associated with $\mathcal{E}$. Consequently, the accumulated noise is diagonalized in the Pauli basis, acting primarily as a stochastic distribution of Pauli errors. This strict simplification of the noise model strictly improves the predictability and reliability of subsequent QEM methods.

    \item \textbf{Zero-Noise Extrapolation (ZNE):} We artificially amplify the device noise by a scale factor $\lambda \ge 1$ using unitary folding. This amounts to substituting $U \to U (U^\dagger U)^n$ such that $\lambda = 2n+1$ \cite{ZNE1, ZNE2}. We then measure the noisy expectation value for any generic
    operator $\mathcal{O}$: $E(\lambda) = \text{Tr}[\mathcal{O} \rho(\lambda)]$ for multiple noise scale factors. We eventually apply a regression model (for instance, linear or exponential) to extrapolate the ideal expectation value at the zero-noise limit:
    \begin{equation}
        E_{\text{mit}} = \lim_{\lambda \to 0} E(\lambda).
    \end{equation}

    \item \textbf{Readout Error Mitigation (TREX):} To mitigate state-preparation and measurement (SPAM) errors, we model the readout noise as a classical transition matrix $\Lambda$. By applying bit-flip twirling prior to measurement, the readout channel is symmetrized into a diagonal effective stochastic matrix \cite{TREX1, TREX2}. The mitigated expectation value of a diagonal observable $\mathcal{O}$ is subsequently recovered by inverting the characterized noise map:
    \begin{equation}
        \langle \mathcal{O} \rangle_{\text{mit}} = \text{Tr}\left[\mathcal{O} \Lambda^{-1}(\rho_{\text{measured}})\right].
    \end{equation}
    This procedure systematically eliminates bias arising from asymmetric assignment errors.
\end{itemize}

\subsection{Conditional resets}
\label{sec:cond_impl}
In stark contrast to the unconditional case, the conditional reset protocol cannot rely on circumventing hardware resets via pure unitary evolution. Because the target reset state depends dynamically on the instantaneous state of the system, according to the transition matrix $P_{i,j}(n)$ in Eq.~(3) of main text, this protocol necessitates active mid-circuit interventions. To achieve this, we designed a specialized hardware algorithm utilizing dynamic circuits and qiskit's classical expressions. We report the pseudocode of the algorithm here below.
\begin{algorithm}[H]    
\begin{algorithmic}[1]
\Require System size $N$, parameters $(J,h)$, trajectory $\mathcal{T}=(a_1,\dots,a_L)$ with $a_\ell\in\{\texttt{u},\texttt{r}\}$
\State Initialize $N$-qubit register in state $\rho$ (as specified in text)
\For{$\ell=1$ to $L$}
    \If{$a_\ell=\texttt{u}$}
        \State Apply unitary $U(J,h,\tau{=}1)$ to all qubits
    \ElsIf{$a_\ell=\texttt{r}$}
        \State Measure all qubits in the computational basis to obtain $m\in\{0,1\}^N$
        \State Reset the register to $\ket{0}^{\otimes N}$
        \State $s \gets \sum_{i=1}^N m_i$
        \If{$s \ge \lceil N/2\rceil$} \Comment{majority of outcomes are $1$}
            \State Apply $\sigma_X^{\otimes N}$ (prepare $\ket{1}^{\otimes N}$)
        \EndIf
    \EndIf
\EndFor
\State Measure all qubits in the computational basis at the end of the circuit and return the final bitstring
\end{algorithmic}
\caption{Conditional reset protocol (majority vote)}
\end{algorithm}
A comprehensive description of the unitary evolution phase is provided in Sec. \ref{sec:Unitary implementation}. The real-time classical evaluation and majority-vote logic (lines 8 and 9) are realized on hardware utilizing the dynamic circuits and classical expression libraries provided by Qiskit \cite{javadiabhari2024quantumcomputingqiskit}. 
\section{Hardware Calibration data}
\label{sec4:hardware_calibration}

All experimental routines were executed on the 127-qubit IBM Quantum processor ibm\_marrakesh. Because superconducting qubit parameters naturally drift over time, we recorded device calibration snapshots concurrently with our experimental execution window from July 15, 2025, to January 27, 2026. Due to the volume of this longitudinal data, the comprehensive calibration history—detailing the relaxation times, dephasing times, and error rates for all utilized physical qubits—is provided as a supplementary dataset. This dataset is accessible via \cite{ZENODO_CITATION}.

To contextualize the longitudinal data recorded on the 127-qubit \texttt{ibm\_marrakesh} processor between July 15, 2025, and January 27, 2026, we monitor several primary calibration metrics. These parameters are critical for understanding the noise environment that the stochastic reset protocols must navigate \cite{nielsen2010quantum}:

\begin{itemize}
    \item \textbf{Relaxation Time ($T_1$):} This represents the characteristic time it takes for a qubit to decay from its excited state $|1\rangle$ to the ground state $|0\rangle$ due to energy loss to the environment. In superconducting circuits, this is often limited by material defects or dielectric loss.
    
    \item \textbf{Dephasing Time ($T_2$):} This measures how long a qubit can maintain a coherent superposition (the relative phase between $|0\rangle$ and $|1\rangle$). $T_2$ is typically shorter than $T_1$ because it is affected by both energy relaxation and pure dephasing caused by magnetic or electrical fluctuations in the hardware environment.
    
    \item \textbf{Single-Qubit Gate Error:} This is the probability that a single-qubit operation, such as the native $RX$ gates used for our transverse field implementation, fails to produce the intended state. These errors are typically on the order of $10^{-4}$ on machines like \texttt{ibm\_marrakesh}.
    
    \item \textbf{Two-Qubit Gate Error:} This refers to the error rate during entangling operations, such as the $CNOT$ or $R_{ZZ}$ gates used for the Ising interactions. These are generally an order of magnitude higher than single-qubit errors because they involve more complex interactions between physical qubits and they are sensitive to crosstalk.
\end{itemize}

\section{Additional results for unconditional resetting}
\label{sec5:additional_unc}
\begin{figure}[H]
    \centering    \includegraphics[width=1.0\textwidth]{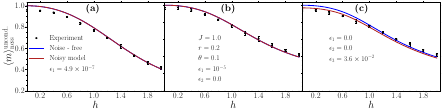}
    \caption{\textit{Comparison between experimental results and theoretical predictions for the unconditional reset protocol.} (a)-(c) Unconditional reset protocol for $N=3$, $5$, and $7$ qubits, respectively, evaluated up to a maximum time $T=10^5$. The plots display the steady-state magnetization $\langle m \rangle_{\mathrm{ness}}^{\mathrm{uncond.}}$ as a function of the transverse field $h$. The experimental hardware data (black dots) are compared with theoretical noise-free predictions (blue solid lines) and the noisy reset-state model (red solid lines). The system parameters are $J=1.0$, $r=0.2$, and $\theta=0.1$. Error bars denote 95\% confidence intervals derived from 5 independent experimental
runs.}
    \label{fig:overview_sm}
\end{figure}

In Fig.~\ref{fig:overview_sm} we provide additional experiments on the unconditional reset protocol. Parameters are analogous to those of the main text except for the smaller value of $\theta=0.1$ used.
We note that the theoretical results match very well the hardware experimental data for all the system sizes displayed. Due to the small $\theta$ value used, the experiment probes a regime close to the actual transverse-field Ising model quantum phase transition. Magnetic ordering is, indeed, monotonically lost and the order parameter decreases as the transverse field $h$ increases. The NESS obtained on the quantum hardware thus reproduces the quantum phase transition of the model in terms of a crossover around $h=1$, where the Floquet spectrum in Eq.~\eqref{eq:bogoliubov_3} becomes gapless. We observe that for unconditional resetting the noiseless unconditional resetting theory in Eq.~\eqref{eq:stationary_state_poissonians} (blue solid lines) already provides a good fit of the hardware data. The effect of noise (red solid lines) is weak, as witnessed by the significantly smaller values of the bitflip probabilities $\epsilon_{1,2,3}$ compared to those obtained for conditional resetting in Fig.~4 of the main text. This effect is caused by the extensive error mitigation used, as detailed in Sec.~\ref{sec:unconditional_impl}, and by the fact that unconditional resetting avoid the usage of dynamic circuits and mid-circuit measurements.

\section{Noise models}
\label{sec6:noise_models}

To characterize the performance of the stochastic reset protocol on the 127-qubit ibm\_marrakesh processor, we analyze several noise models to account for hardware imperfections. We utilize numerical minimization via the COBYQA algorithm to optimize the parameters for each model. We minimize the mean squared error (MSE) between the theoretical steady-state predictions and the experimental hardware results. These models range from simple single-qubit decoherence to complex correlated errors that mimic the noisy interactions inherent in the superconducting heavy-hex lattice. In all these cases, we describe noise in the theory by considering a reset-free dynamics which is not purely unitary, as described in Eq.~\eqref{eq:unitary_gate}, but in the form of a Kraus evolution. The density matrix $\rho_{n+1}$ at discrete step $n+1$ is thus given in terms of that at step $n$ by
\begin{equation}
\rho_{n+1}=\sum_{\vec{\mu}}K_{\vec{\mu}} \, \rho_n \, K_{\vec{\mu}}^{\dagger}, 
\label{eq:kraus}
\end{equation}
where the Kraus operators $K_{\vec{\mu}}$ acting on $N$ qubits $\vec{\mu}=(\mu_1,\mu_2,\dots \mu_N)$ satisfy the completeness relation $\sum_{\vec{\mu}}K_{\vec{\mu}}^{\dagger}K_{\vec{\mu}}=\mathbb{I}$. We consider for Kraus operators the following form
\begin{equation}
K_{\vec{\mu}}=K_{(\mu_1,\mu_2\dots \mu_N)}=(K_{\mu_1} \otimes K_{\mu_2}\dots \otimes K_{\mu_N}) U_{\theta},
\label{eq:Kraus_single_body}
\end{equation}
with $K_{\mu_i}$ single-qubit operators, $\otimes$ denoting tensor product, and $U_{\theta}$ the many-body unitary gate as in Eqs.~\eqref{eq:ising_discrete_unitary_sm} and \eqref{eq:ising_discrete_terms_sm}.
We now define the Kraus operators associated to each noise model. 

\subsection*{Depolarizing Channel (DP)}
The depolarizing channel represents a symmetric error where a qubit is replaced by the maximally mixed state with probability $p$. For each qubit $\mu$, there are four possible Kraus operators $K_{\mu}=\{K_0,K_1,K_2,K_3\}$ in Eq.~\eqref{eq:Kraus_single_body}. These are given by:
\begin{equation}
K_0 = \sqrt{1-p}I, \quad K_1 = \sqrt{\frac{p}{3}}\sigma_x, \quad K_2 = \sqrt{\frac{p}{3}}\sigma_y, \quad K_3 = \sqrt{\frac{p}{3}}\sigma_z,
\label{eq:depolarization}
\end{equation}
This model serves as a general benchmark for isotropic noise across the $N=3, 5, \text{ and } 7$ system sizes studied.

\subsection*{Phase Damping (PD)}
Phase damping (or dephasing) describes the loss of quantum coherence without the loss of energy, a common occurrence in superconducting qubits during the experimental window. For each qubit we have two possible Kraus operators $K_{\mu}=\{K_0,K_1\}$ with
\begin{equation}
K_0 = \begin{pmatrix} 1 & 0 \\ 0 & \sqrt{1-\lambda} \end{pmatrix}, \quad K_1 = \begin{pmatrix} 0 & 0 \\ 0 & \sqrt{\lambda} \end{pmatrix},
\label{eq:phase_damping}
\end{equation}
where $\lambda$ relates to the dephasing time $T_2$.

\subsection*{Generalized Amplitude Damping (AD)}
The generalized amplitude damping channel models the thermal relaxation of the qubits due to coupling to an environment at finite temperature \cite{nielsen2010quantum}. For each site we have the following four Kraus operators:
\begin{align}
K_0 = \sqrt{p} \begin{pmatrix} 1 & 0 \\ 0 & \sqrt{1-\gamma} \end{pmatrix}, \quad & K_1 = \sqrt{p} \begin{pmatrix} 0 & \sqrt{\gamma} \\ 0 & 0 \end{pmatrix} \\
K_2 = \sqrt{1-p} \begin{pmatrix} \sqrt{1-\gamma} & 0 \\ 0 & 1 \end{pmatrix}, \quad & K_3 = \sqrt{1-p} \begin{pmatrix} 0 & 0 \\ \sqrt{\gamma} & 0 \end{pmatrix},
\label{eq:amplitude_damping}
\end{align}
where $\gamma$ is the probability of energy exchange and $p$ is related to the temperature of the environment ($p=1$ corresponding to the stationary state $\ket{0}$ and thus to a zero-temperature environment). This process describes the $T_1$ relaxation dynamics. 

\subsection*{Correlated ZZ Error (ZZ)}
Given the heavy-hex connectivity and the use of SWAP gates for periodic boundary conditions, we include a correlated two-qubit dephasing channel. This model assumes that with probability $p_{zz}$, a $Z \otimes Z$ error occurs between adjacent qubits $i$ and $i+1$. In this case, the decomposition in Eq.~\eqref{eq:kraus} is performed in terms of Kraus operators $K_{i,i+1}$ acting on neighboring qubits. For the latter we have two possibilities $K_{i,i+1}=\{K_0,K_1\}$ as
\begin{equation}
K_0 = \sqrt{1-p_{zz}} (I \otimes I), \quad K_1 = \sqrt{p_{zz}} (\sigma_z^{(i)} \otimes \sigma_z^{(i+1)}),
\label{eq:zz_noise}
\end{equation}
This channel is applied sequentially across the 1D chain to mimic noisy $ZZ$ interactions often found in superconducting architectures.

\subsection{Comparison between different noise models for conditional resetting}
In table \ref{tab:noise}, we report the mean squared error for the optimized values of the parameters characterizing each noise model. On the first row, we report the noisy reset-state model (``Reset state error'') described in the main text. 
\begin{table}[htpb]
\centering
\label{tab:noise_model_accuracy}
\begin{tabular}{lccc}
\hline\hline
\textbf{Noise Model} & \textbf{$N=3$} & \textbf{$N=5$} & \textbf{$N=7$} \\
\hline
Reset state error & $\mathbf{1.5 \times 10^{-5}}$ & $\mathbf{1.2 \times 10^{-4}}$ & $5.7 \times 10^{-4}$ \\
DP & $2.4 \times 10^{-4}$ & $1.1 \times 10^{-3}$ & $5.95 \times 10^{-4}$ \\
DP \& PD & $1.8 \times 10^{-4}$ & $1.07 \times 10^{-3}$ & $6.5 \times 10^{-4}$ \\
DP \& PD \& AD & $1.7 \times 10^{-4}$ & $1.0 \times 10^{-3}$ & $6.5\times10^{-4}$ \\
DP \& PD \& AD \& ZZ & $1.7 \times 10^{-4}$ & $1.05 \times 10^{-3}$ & $\mathbf{3.9 \times 10^{-4}}$ \\
\hline\hline
\end{tabular}
\caption{Comparison of noise model accuracies across different system sizes ($N=3, 5, 7$). The values presented represent the mean squared error (MSE) between the steady-state theoretical predictions incorporating the respective noise model and the experimental hardware results. The optimal parameters for each model were obtained via numerical minimization using the COBYQA algorithm.}
\label{tab:noise}
\end{table}

\begin{figure}[t]
    \centering
    \includegraphics[width=1.01\textwidth]{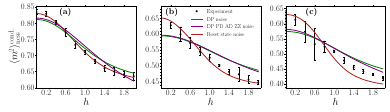}
    \caption{\textit{Comparison of various noise models with experimental hardware data for conditional resetting protocol.} (a)-(c) Conditional reset protocol for $N=3,5,$ and $7$ qubits, respectively, evaluated up to a maximum time $T=400$. The panels display the steady-state expectation value $\langle m^2 \rangle_{\mathrm{ness}}^{\mathrm{cond.}}$ as a function of the transverse field $h$. The experimental hardware data (black dots) are contrasted with theoretical trajectories incorporating depolarizing model (DP noise), a composite model of identified noise sources (DP, PD, AD, and ZZ noise), and the noisy reset-state model described in the main text. The color of the solid line corresponding to each noise model is reported in the legend of panel (b). Error bars denote 95\% confidence intervals derived from 5 independent experimental
runs.}
\label{fig:noise_comparison}
\end{figure}

In Fig.~\ref{fig:noise_comparison}, we show the comparison between various noise models and the hardware data. We contrast the performance of the depolarizing (DP) model in Eq.~\eqref{eq:depolarization}, the composite model incorporating all identified noise sources (DP, PD, AD, and ZZ) in Eqs.~\eqref{eq:depolarization}-\eqref{eq:zz_noise}, and the noisy reset-state error model discussed in the main text. From Table \ref{tab:noise} one can see that for $N=3$ and $N=5$ the noisy reset-state model provides a mean-squared error that is one order of magnitude less than that of all the other noise models. The noisy reset-state model manifestly outperforms the other models, as one can see from Fig.~\ref{fig:noise_comparison}(a) and (b). For $N=7$ in Fig.~\ref{fig:noise_comparison}(c), the values of the mean squared error of the various models are of the same order of magnitude. Only the composite noise model (DP, PD, AD and ZZ) yields a mean squared error slightly smaller than that of the noisy reset-state model. The latter contains, however, only three fitting parameters, rather than five as in the other model. For this reason, and given that the noisy reset-state model admits a neat physical interpretation owing to the $\mathbb{Z}_2$ symmetry of the Ising model, we adopted this noise model to compare with the hardware experimental data in the main text.  

\end{document}